\documentclass[aps,pra,twocolumn,showpacs]{revtex4-1}
\usepackage{graphicx}
\usepackage{amsmath}
\usepackage{amssymb}
\usepackage{color}
\usepackage{float}

\usepackage[colorlinks=true, citecolor=blue, urlcolor=blue ]{hyperref}

\input{epsf}
\begin{document}

\title{Adiabatic freezing of entanglement with insertion of defects\\ in one-dimensional Hubbard model}

\author{Sreetama Das\(^{1}\), Sudipto Singha Roy\(^{1,2}\), Himadri Shekhar Dhar\(^{1,3}\), 
Debraj Rakshit\(^{1,4}\),\\ Aditi Sen(De)\(^{1}\), and Ujjwal Sen\(^{1}\)}

\affiliation{\(^1\)Harish-Chandra Research Institute, HBNI, Chhatnag Road, Jhunsi, Allahabad 211 019, India\\
\(^2\)Department of Applied Mathematics, Hanyang University (ERICA), 55 Hanyangdaehak-ro, Ansan, Gyeonggi-do, 426-791, Korea\\
\(^3\)Institute for Theoretical Physics, Vienna University of Technology, Wiedner Hauptstra{\ss}e 8-10/136, 1040 Vienna, Austria\\
\(^4\)Institute of Physics, Polish Academy of Sciences, Aleja Lotnik{\'o}w 32/46, PL-02668 Warsaw, Poland}


\begin{abstract}
We report on ground state phases of a doped one-dimensional Hubbard model, which for large onsite interactions is governed by the $t$-$J$ Hamiltonian, where the extant entanglement is immutable under perturbative or sudden changes of system parameters, a phenomenon termed as adiabatic freezing. We observe that in the metallic Luttinger liquid phase of the model bipartite entanglement decays polynomially and is adiabatically frozen, in contrast to the variable, exponential decay in the phase-separation and superconducting spin-gap phases. Significantly, at low fixed electron densities, the spin-gap phase shows remarkable affinity to doped resonating valence bond gas, with multipartite entanglement frozen across all parameter space. 
We note that entanglement, in general, is sensitive to external perturbation, as observed in several systems, and hitherto, no such invariance or freezing behavior has been reported.
%
\end{abstract}

\maketitle

\section{Introduction}
Over the years, a challenging task has been to explore how entanglement~\cite{horo} is distributed among the constituents of a many-body system and understand its 
effects on cooperative phenomena~\cite{amico,adv,mac}.
For instance, it was observed that the constituents of the non-critical phases of many-body systems are, in general, less entangled with particles beyond their 
nearest neighbors (NN), and  obey the area law of scaling of entanglement entropy~\cite{hastings,plenio}, which provides useful information about 
their ground state properties~\cite{amico,adv,mac,plenio} and is closely related to its numerical simulability~\cite{dmrg,mps}. 
Hence, study of quantum correlation may actually provide deeper insight about the underlying cooperative and critical phenomena in these systems~\cite{sachdev1,sondhi,vojta}. 
In return, quantum many-body systems are also important substrates for quantum communication \cite{comm, q_communication1}
and computation  protocols \cite{comp,kit}, and are thus key enablers for quantum technology. 

In this work, we report on the entanglement behavior in the ground state phases of a doped one-dimensional (1D) Hubbard model with large onsite interactions. The quantum spin-1/2 particles on the lattice doped with holes interact via the  $t$-$J$ Hamiltonian \cite{t_j}, 
with \(t\) representing a typical tunneling strength between two neighboring sites and \(J\) serving as the spin-spin interaction strength between 
particles in filled neighboring sites. 
The $t$-$J$ Hamiltonian is widely used to study the physical properties of doped quantum spin systems, in particular for high-$T_c$ superconducting phases of 
strongly-correlated matter \cite{t-J_sc,wen}. The minimum energy configuration of the $t$-$J$ Hamiltonian exhibits a rich phase diagram in the $J/t$-$n_{el}$ plane, with $n_{el}$ being the electron concentration or density, and has already been extensively studied using physical quantities such as ground state energy, spin correlation functions, and spin gap\cite{phase_separation,t-j_phase1,t_j1_j2,nakamura,t-j_phase2}. In this regard, one of our primary motivations is to investigate how quantum correlations, especially bipartite entanglement (BE) and multipartite entanglement (ME), behave in these different phases, and whether insertion of defects play a significant role in altering the entanglement properties. 

The key finding of this work is the existence of entanglement in the ground state of the doped 1D $t$-$J$ Hamiltonian, in particular at low electron densities, which remains invariant under sudden or perturbative changes to the  $J/t$ ratio, 
implying potential application in robust quantum technologies~\cite{kara-kara-ekhane}. In other words, the entanglement remains constant under perturbations of the system parameter, a phenomena reminiscent of the \emph{adiabatic freezing} of quantum correlations~\cite{hsd-freeze} (cf.~\cite{manis,expt,freezing}), where the aforementioned 
quantities are completely insensitive or \emph{frozen} with respect to changes in system parameters~\cite{hsd-freeze} or decoherence~\cite{manis}. We observe that this adiabatic freezing behavior of entanglement is different for bipartite and multipartite cases, and is closely related to the relevant ground state phases of this model~\cite{phase_separation,t-j_phase1,t_j1_j2,nakamura,t-j_phase2}.  
To elaborate, we observe that at low $J/t$ ratio ($J/t < 2$), for low $n_{el}$, when the  system is known to lie in the metallic Luttinger liquid phase \cite{t-j_phase1}, two-site BE, as quantified by the logarithmic negativity \cite{adotey-negativity,negativity1}, decays polynomially with the 
increase in lattice distance, $r$ = $|i-j|$, between the lattice sites \(i\) and \(j\), which essentially signals the dominating long-range order in the phase. Interestingly, within the metallic phase, the BE is invariant to changes in the 
$J/t$ ratio and is therefore adiabatically frozen. In contrast, at higher $J/t$ ratio, superconducting spin-gap phase~\cite{t_j1_j2,nakamura} and electron-hole phase separation (PS) occurs~\cite{phase_separation}, accompanied by an exponential decay of BE. Subsequently, the adiabatic freezing of BE is lost during the quantum phase transition.
Of greater significance is the behavior of multipartite entanglement, which for low fixed values of $n_{el}$, remains adiabatically frozen for all values of the $J/t$ parameter space.
Using generalized geometric measure (GGM) \cite{ggm} (cf.~\cite{ggm_extra}) as the measure of genuine 
multipartite entanglement, we show that the variation of GGM across the $J/t$-$n_{el}$ phase space, for low $n_{el}$, remains invariant under adiabatic changes of the  $J/t$ ratio. It is important to note 
that no such adiabatic freezing of ME is observed in the undoped anisotropic 1D model  \cite{diverging-roy}. 
Rather counterintuitively, it appears that 
the presence of \emph{impurities} or \emph{defects} (as modeled by the holes)
in the spin chain acts as a vehicle for phases with \emph{frozen} ME. The importance of the results lie in the fact that many-body systems with robust ME, which is not sensitive to perturbations in system parameters or environmental processes, are necessary for realizing quantum information-theoretic protocols such as measurement based quantum computation \cite{comp} and quantum communication protocols \cite{comm, q_communication1}. The paper is arranged as follows. In Sec.~\ref{model} we introduce the 1D $t$-$J$ Hamiltonian. We study the decay and adiabatic freezing of bipartite entanglement in Sec.~\ref{be}. We discuss the low electron density ground states of the model in Sec.~\ref{gs} and demonstrate the freezing of genuine multipartite entanglement in Sec.~\ref{me}. We conclude in Sec.~\ref{conc}.

\section{\label{model}Model}
In our study, we consider the $t$-$J$ Hamiltonian as the 
structure that governs the interaction 
between the quantum particles in the doped 1D spin lattice, with $N$ sites populated with $N_{el} (< N)$ quantum spin-1/2 particles. The rest of the sites are vacant or contain \emph{holes}. The ``electron density'' of the lattice is given by $n_{el}$ ($= N_{el}/N$).
The $t$-$J$ Hamiltonian 
can be obtained perturbatively from the prominent Hubbard model in the limit of large on-site interaction \cite{t_j}, 
and has been expressed in literature in the form,
\begin{equation}
H= -t\sum_{\langle i,j\rangle,\sigma} {\mathcal{P}_G}~ (c_{i\sigma}^{\dagger}c_{j\sigma}+\text{h.c.}) ~{\mathcal{P}_G}+J\sum_{\langle i,j\rangle}\vec{S_i}\cdot\vec{S_j},
\label{tJ}
\end{equation}
where $c_{i\sigma}$ ($c_{i\sigma}^\dag$) is the fermionic annihilation (creation) operator of spin $\sigma$ ($= \{\uparrow, \downarrow\}$), acting on site $i$. $\mathcal{P}_G$ is the Gutzwiller projector $\Pi_i (1-n_{i\uparrow}n_{i\downarrow})$ which enforces at most single occupancy at each lattice site.
$S_i = \frac{1}{2}\sigma_i$'s are the triad of spin operators $\{S^x,S^y,S^z\}$, while $t$ and $J$ correspond to the transfer energy and the spin-exchange interaction energy terms, 
respectively, and each is limited to nearest-neighbor sites, with periodic 
boundary condition. The ground state phase diagram for the above 1D model has received widespread attention in the past years 
\cite{phase_separation,t-j_phase1,t_j1_j2,nakamura,t-j_phase2}. In particular, the presence of three primary phases, namely the repulsive Luttinger liquid or metallic, 
attractive Luttinger liquid or superconducting, and the phase separation, have been predicted using exact diagonalization \cite{t-j_phase1}. 
However, recent results, using density matrix renormalization group techniques, have also reported 
the presence of a superconducting spin-gap phase at low $n_{el}$ \cite{t-j_phase2, nakamura}. These phases play a significant role in the entanglement properties of the doped quantum spin model.

\section{Decay of bipartite entanglement and adiabatic freezing in metallic phase} 
\label{be} 
We now focus on the behavior of bipartite entanglement in the ground state of the $t$-$J$ Hamiltonian. In particular, we look at the logarithmic negativity ($\mathcal{E}$) in the state, $\rho_{ij}$, shared between two-sites \(i\) and \(j\), and its decay with increase in lattice distance, $r$ = $|i-j|$, for different phases of the model in the $J/t$-$n_{el}$ plane.
\begin{figure}[t]
\includegraphics[width=0.42\textwidth]{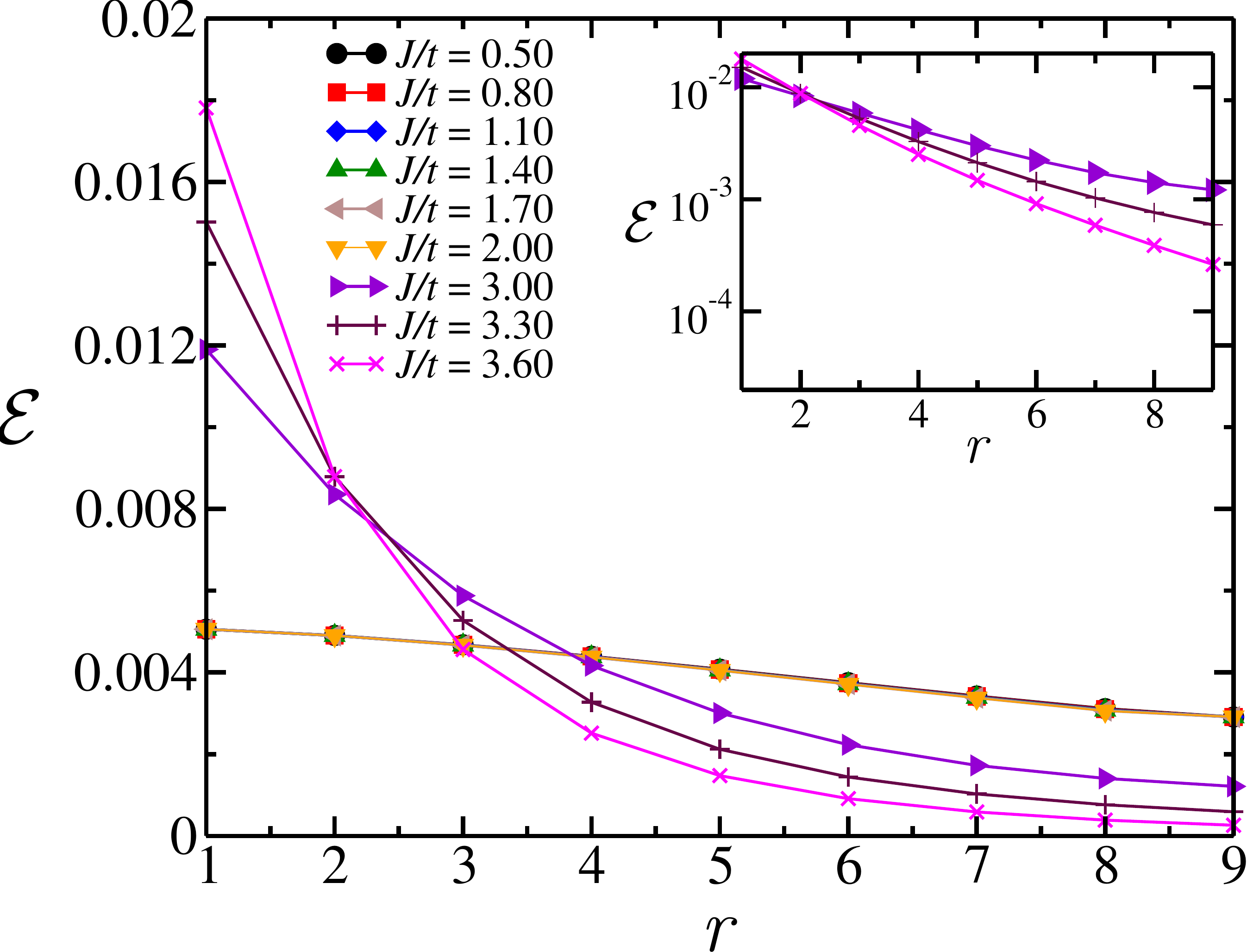}
\caption{(Color online.) Decay and adiabatic freezing of bipartite entanglement in phases of the $t$-$J$ Hamiltonian. 
The plot shows the variation of two-site entanglement ($\mathcal{E}$) with increase in lattice distance $r$ = $|i-j|$,  for the 1D $t$-$J$ Hamiltonian, 
with $N = 30$ and $n_{el}=\frac{2}{N}$. 
For $J/t \leq 2$, the ground state remains in the metallic phase and $\mathcal{E}$ decays polynomially as $1/(A + Br)$, with $r$, exhibiting the presence of a dominating long-range order in the ground state. The values of $A=162.6$ and $B=18.9$, obtained from the average best-fitted curve, remains almost unchanged for all the curves in this phase, and BE is adiabatically frozen. This  freezing behavior of bipartite entanglement is shown more clearly in Fig. \ref{BE_freezing}. In contrast, for $J/t \geq 3$, 
the superconducting and PS phases leads to exponential decay of BE, given by $\mathcal{E} \sim C ~ \exp{(-\frac{r}{\xi})}$, 
where $\xi$ is the characteristic length and the constant $C$ can be obtained from the best-fitted curve. $\xi$ and $C$ are dependent on $J/t$ and the adiabatic freezing of BE is lost in this phase.  The vertical axis are in ebits and the horizontal axes are dimensionless. \(J/t\) is also dimensionless.
 In the inset, we  set the vertical axis in the logarithmic scale and plot $\mathcal{E}$ for $J/t\geq 3$. }
\label{fig1}
\end{figure}
For a bipartite state $\rho_{ij}$, shared between two sites \(i\) and \(j\), its 
 logarithmic negativity is defined as
\begin{eqnarray}
\mathcal{E}(\rho_{ij})=\text{log}_2(2 \mathcal{N}(\rho_{ij})+1),
\end{eqnarray}\\
where $\mathcal{N}$ is the negativity \cite{adotey-negativity,negativity1},  defined as the absolute value of the sum of the negative eigenvalues of $\rho^{T_i}_{ij}$,  so that
$\mathcal{N}(\rho_{ij})=\frac{||\rho^{T_i}_{ij}||_1-1}{2}$, where $\rho^{T_i}_{ij}$ denotes the partial transpose of $\rho_{ij}$ with respect to the subsystem $i$.

\begin{figure}[t]
\includegraphics[width=0.42\textwidth]{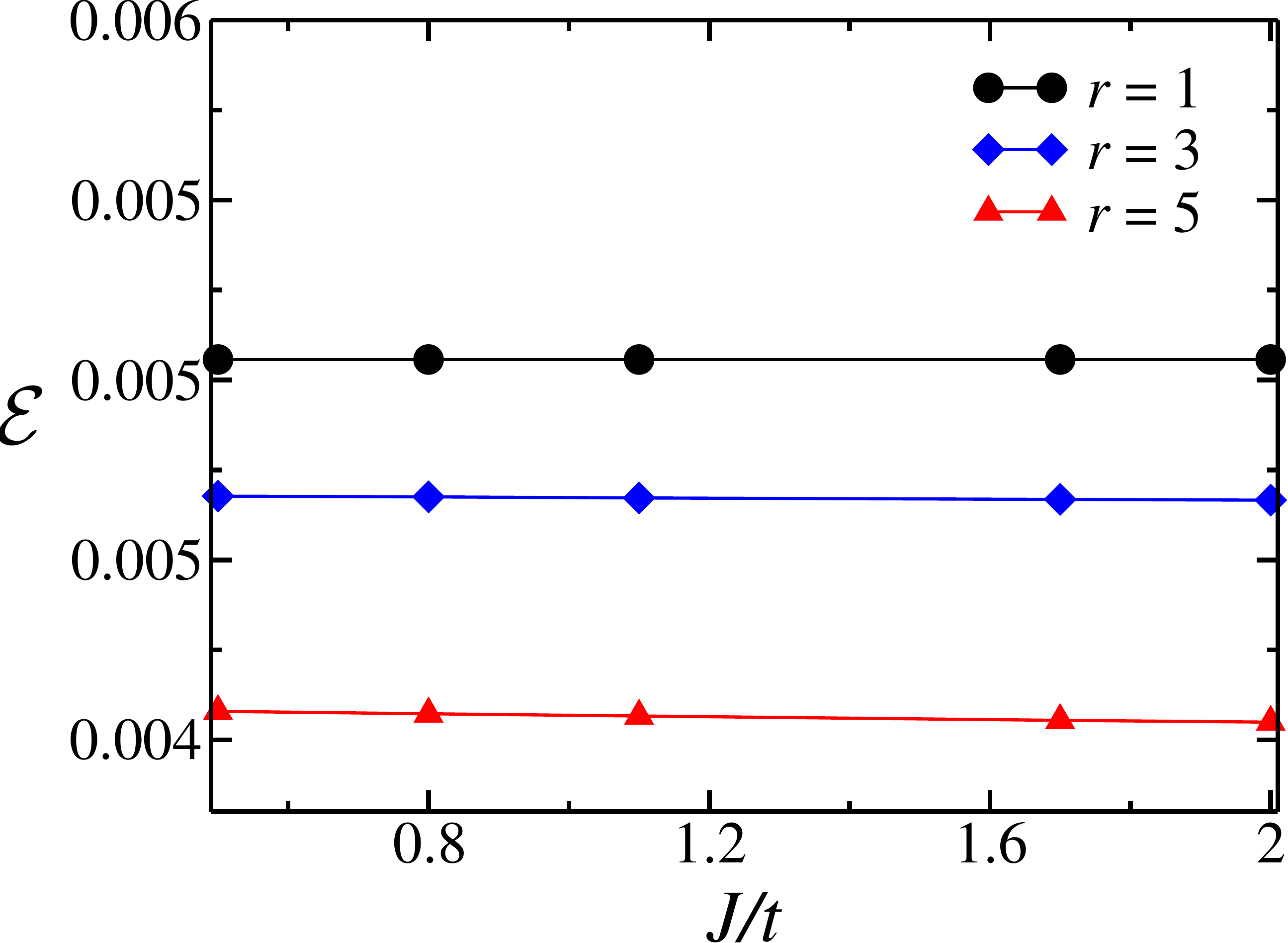}
\caption{(Color online.)
Adiabatic freezing of bipartite entanglement. Variation of two-site entanglement ($\mathcal{E}$) with $J/t$ for different  lattice distances, $r$ = $|i-j|$=1 (black-circle), $r =3$ (blue-diamond), $r=5$ (red-triangle), for the 1D $t$-$J$ Hamiltonian, 
with $N = 30$ and $n_{el}=\frac{2}{N}$. 
From the figure, one can see that for $J/t \leq 2$, the ground state BE remains  adiabatically frozen.  The vertical axis is in ebits and the horizontal axis is dimensionless. Although the region considered in the figure is $ 0.5 \leq J/t \leq 2 $, the freezing behaviour extends all the way to $ J/t=0 $. }
\label{BE_freezing}
\end{figure}
The decay of spin correlation functions with inter-site distance $r$, often signals the nature of correlation present in the system 
\cite{sachdev1,sondhi,class_corr_phase}. In general, for non-critical states of strongly-correlated 1D spin systems, quantum correlations are 
short-ranged and decay exponentially with the increase of lattice distance \cite{lieb_robinson}, giving rise to features such as the area law 
\cite{hastings,plenio}.
%
As discussed earlier, for all $n_{el}$ in the $J/t$-$n_{el}$ phase space, at low values of $J/t$ ($\approx 2$), the ground state remains in a metallic phase or a repulsive 
Luttinger liquid-like phase \cite{t-j_phase1}. 
In Fig.~\ref{fig1}, we plot the decay of bipartite entanglement, $\mathcal{E}(\rho_{ij})$, with the lattice distance $r$, 
for different values of 
the $J/t$ ratio, using exact diagonalization to obtain the ground state for 
$N = 30$  and $n_{el}$ = ${2}/{N}$ \cite{lanczos}. 
In the metallic phase (\(J/t \leq 2.0\)), the decay with respect to \(r\) can be encapsulated as  \(\mathcal{E} \sim 1/(Ar+B)\), 
where the numerically obtained  values of  $A$ and $B$, from the best-fit curve, are given {by $A = 162.6$ and $B = 18.9$,} respectively.
Significantly, the curves of \(\mathcal{E}(\rho_{ij})\) with respect to \(r\) for different values of \(J/t\) are almost invariant in the metallic phase, i.e., the decay is not only polynomial, but it is the same polynomial for all \(J/t\) (see Fig.~\ref{BE_freezing} for a more clear illustration).
 The entanglement therefore remains adiabatically frozen under perturbations of $J/t$.
It is known that in the Luttinger-liquid phase, the NN spin correlation functions are independent of $J/t$ and the electron density\cite{t-j_phase2}. 
Therefore, one can infer that the freezing of bipartite entanglement is characteristic of the ground state phase diagram of the 1D $t$-$J$ model. However, for non-NN spin correlation functions there is a very slow variation with the system parameters.
Therefore, the behavior of 
$\mathcal{E}$ in Fig.~\ref{fig1} not only expectedly follows the properties of spin correlation functions but also provides more insight about the ground state in the metallic phase. 
The freezing of bipartite entanglement with respect to system parameters can be advantageous for implementing quantum technologies that is robust to fluctuations in the system parameters, potentially due to errors in the preparation procedure~\cite{kara-kara-ekhane}. 

In Fig.~\ref{fig1}, for higher values of $J/t$ ($\geq 3$), when the system subsequently  enters into the superconducting and phase-separation region \cite{t-j_phase1,t-j_phase2}, the ground state of the system is likely to be a spin liquid or superposition of the terms where all the spin-1/2 particles form clusters, leading to a distinctive electron-rich and hole-rich phase separation, 
respectively. 
Consequently, in these regions, spin correlation functions are likely to be short-ranged similar to undoped ground state of the Heisenberg model. In other words, for high $J/t$, an exponential decay of spin correlation functions is expected. 
From Fig.~{\ref{fig1}}, it is quite prominent that as the $J/t$ ratio increases, the BE measure $\mathcal{E}(\rho_{ij})$ 
exhibits an exponential decay with the increase of $r$, given by $\mathcal{E}\sim C~ \exp({-{r}/{\xi}})$, where $\xi$ is the characteristic length of the decay. 
Again from the best-fit data, one can estimate the value of the constant $C$. As an example, for $J/t$ = 3.6, the best-fitted plot yields $C= 0.0236$ and $\xi=0.5225$.
Interestingly, in contrast to the polynomial decay of BE in the metallic phase, the exponential decay rate is not constant for different values of $J/t$ in 
the superconducting and PS phase. It is observed that the decay becomes steeper, with increase in $J/t$, such that entanglement vanishes quicker with \(r\), and the freezing behavior is completely lost in these regions.
 
Moreover, if we introduce additional next-nearest neighbor interactions in the 
$t$-$J$ Hamiltonian, 
the subsequent spin model is known to have a rich phase diagram in the $J/t$-$n_{el}$ plane \cite{t_j1_j2}, which is qualitatively similar to that 
of the Hamiltonian in Eq.~\ref{tJ}, apart from the fact that, in this case, the intermediate spin-gap phase is spread over 
a larger area in the phase plane. The boundaries between the metallic, superconducting, and PS phases are altered. Interestingly, the freezing of BE, or lack thereof, in the different phases remains unaltered.

\begin{figure}[h]
\includegraphics[width=0.36\textwidth]{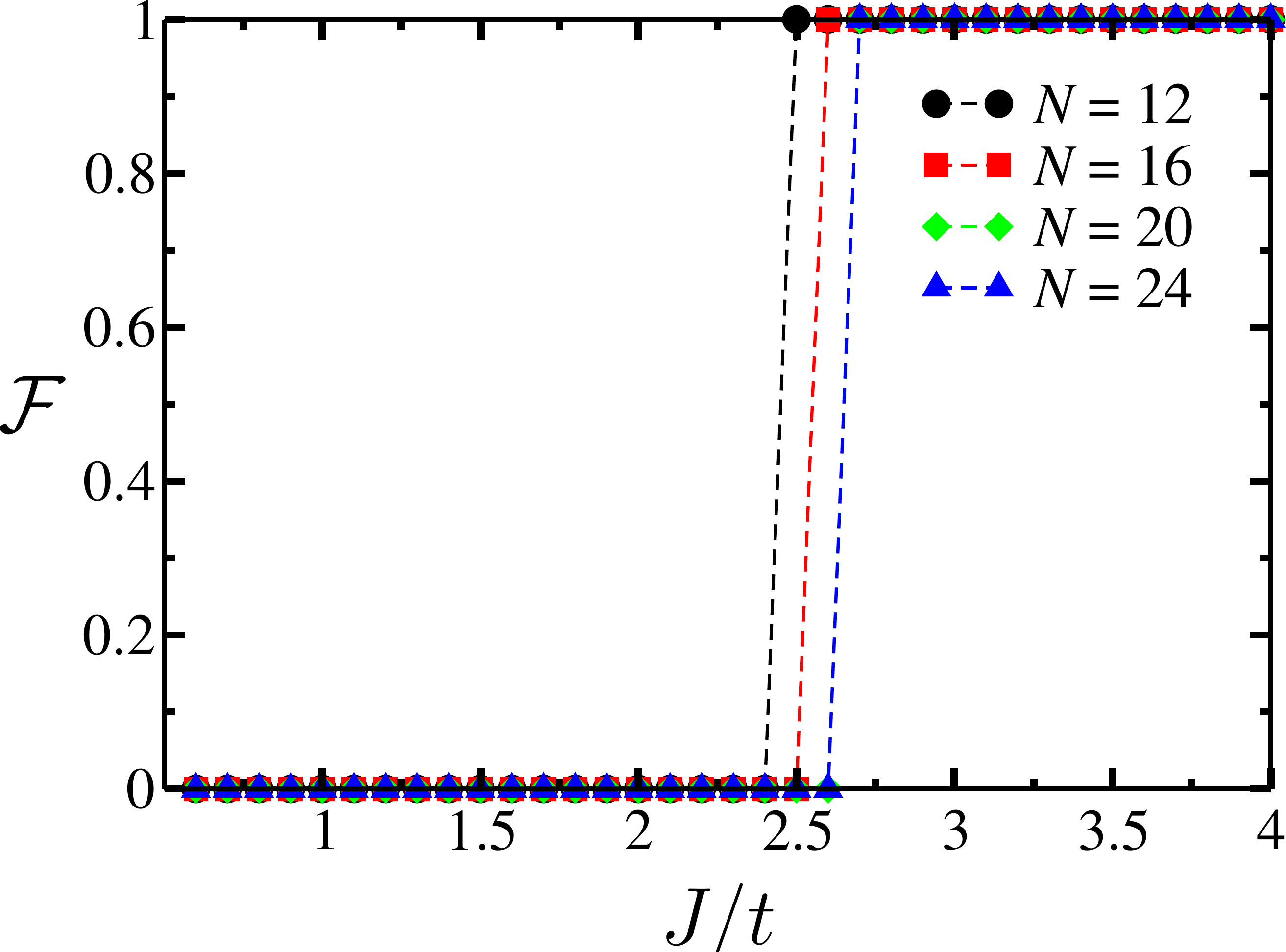}
\caption{(Color online.) RVB gas as the spin-gap phase of the \(t\)-\(J\) Hamiltonian at low electron densities. We plot the fidelity ($\mathcal{F}$) of the ground state of the 1D $t$-$J$ Hamiltonian, obtained via exact diagonalization, and the variational long-range RVB state, at electron density $n_{el}=2/N$. The curves shown in the figure pertain to 
1D lattices with $N = 12, 16, 20, 24$ sites. Note that the curves corresponding to \(N \geq 24\) coincide with reasonable numerical accuracy. 
We note that the RVB gas state considered for different values of \(J/t\) and \(N\) are not the same, as the set \(\{r_{\mathcal{C}}\}\) that maximizes the fidelity are different.
All quantities used are dimensionless. }
\label{fidelity}
\label{LN}
\end{figure}

\section{\label{gs}Ground state phase at low electron densities}
 To understand the behavior of bipartite entanglement in the different phases of the 1D $t$-$J$ Hamiltonian,
we now discuss the ground state properties of the  model at low electron densities. In the superconducting phase of the model, 
at low $n_{el}$, a finite spin gap opens up, which is in contrast to the behavior at the high density region where the system remains gapless~\cite{t-j_phase2}. Interestingly, we find that in this spin-gap phase, the ground state of the system is essentially a long-range resonating valence bond (RVB) state or the RVB gas \cite{RVB_gas}. Thus, the ground state can be expressed as 
\begin{equation}
|\psi\rangle_{\texttt{RVB}} = \sum_{\mathcal{C}}  r_{\mathcal{C}}~ \prod_{i\neq j} |A_i B_{j}\rangle \otimes \prod_{k} |0_k\rangle,
\end{equation}
where $|A_i B_j\rangle$ = $\frac{1}{\sqrt{2}}(|1\rangle_i |2\rangle_j - |2\rangle_i |1\rangle_j)$  is the spin singlet formed between two spin-1/2 particles at spin-occupied sites `$i$' and `$j$', corresponding to the sublattices $A$ and $B$, respectively. 
The product is over all such non-overlapping dimers between $N_{el}/2$ pairs of spin-occupied sites $\{i,j\}$. The state $\prod_{k} |0_k\rangle$ 
represents the $k$ holes at $N - N_{el}$ vacant sites. The summation  corresponds to the superposition of all possible dimer coverings ($\mathcal{C}$) 
on the lattice, each with relative weight $r_{\mathcal{C}}$. 

The RVB gas description of the superconducting spin-gap phase of the 1D $t$-$J$ Hamiltonian, at low electron density, 
has a remarkable significance, since it allows for the study related to the phase properties of this
model and beyond, using
the RVB ansatz \cite{t_j1_j2,t_J_RVB2,t_J_RVB3} under suitable doping. 
Hence, even for moderate-sized systems, where  exact diagonalization is not possible, the doped RVB ansatz opens up the possibility of investigating different properties of 
the $t$-$J$ Hamiltonian~\cite{t-J_RVB_ours} using tensor network~\cite{cirac} or other approximate approaches~\cite{isotropic}. Fig.~\ref{fidelity} depicts the behavior of the fidelity,
$\mathcal{F}$ = $\max_{\{r_{\mathcal{C}}\}}|\langle \phi_g|\psi\rangle_{RVB}|$, between the 
ground state \(|\phi_g\rangle\) as estimated by exact diagonalization and the RVB state 
\(|\psi\rangle_{RVB}\),
for low electron density, $n_{el} = 2/N$. One observes that after a certain $J/t$ 
($\approx 2.3$), pertaining to the transition between the metallic and superconducting phases, the minimum energy configuration of the system is actually long-ranged RVB gas.
Further observation shows that even if we increase the $J/t$ ratio to a large value, the ground state at low $n_{el}$
still exhibits RVB behavior but the 
probability of formation of nearest-neighbor singlet pairing increases as compared to distant pairs due to the formation of electron-hole phase separation. In principle, 
this may lead to the formation of an RVB liquid state or NN dimer phase for high \(J/t\),
which has a decisive bearing on the exponential decay pattern of the two-site entanglement of the system as   the quantum correlation of the NN RVB states are known to be short-ranged.

\section{\label{me}freezing of multipartite entanglement}
A significant outcome of our analysis of the entanglement properties of ground state phases of the 1D $t$-$J$ Hamiltonian, is the existent characteristics of genuine multipartite entanglement. To measure the genuine ME in the different
regions of the $J/t$-$n_{el}$ plane, we use the generalized geometric measure (GGM)\cite{ggm} (cf.~\cite{ggm_extra}). 

For an $N$-party pure quantum state $|\phi\rangle$, the GGM is a computable measure of genuine multisite entanglement, which is formally defined as the optimized 
fidelity-based distance of the state from the set of all states that are not genuinely multiparty entangled. Mathematically, the GGM can be evaluated as
\begin{eqnarray}
\mathcal{G}(|\phi\rangle) = 1 - \lambda_{\text{max}}^2(|\xi_N\rangle) \nonumber,
\end{eqnarray}
where $\lambda_{\text{max}}$ =  $\max|\langle\xi_N|\phi\rangle|$, and $|\xi_N\rangle$ is an $N$-party 
non-genuinely multisite entangled quantum state and the maximization is performed over the set of all such states. 
The GGM can be effectively computed using the 
relation
\begin{eqnarray}
\mathcal{G}(|\phi\rangle) = 1 - \text{max}\{\lambda^2_{A:B}|A\cup B = {A_1 , \dots, A_N }, A\cap B=\phi\}, \nonumber
\end{eqnarray}
where $\lambda_{A:B}$ is the maximum Schmidt coefficient in all possible bipartite splits $A:B$ of the given state $|\phi\rangle$. 
A complexity in computation of the multiparty entanglement measure $\mathcal{G}$ lies in the fact that the number of possible bipartitions 
 increases exponentially with an increase of the lattice size. 
Therefore, we need to restrict ourselves to moderate-sized systems only, which in our case 
restricts us
to $N = 16$.
We observe that at low electron concentrations the GGM is adiabatically frozen over significant regions of the phase space.

\begin{figure}[t]
\includegraphics[width=0.35\textwidth]{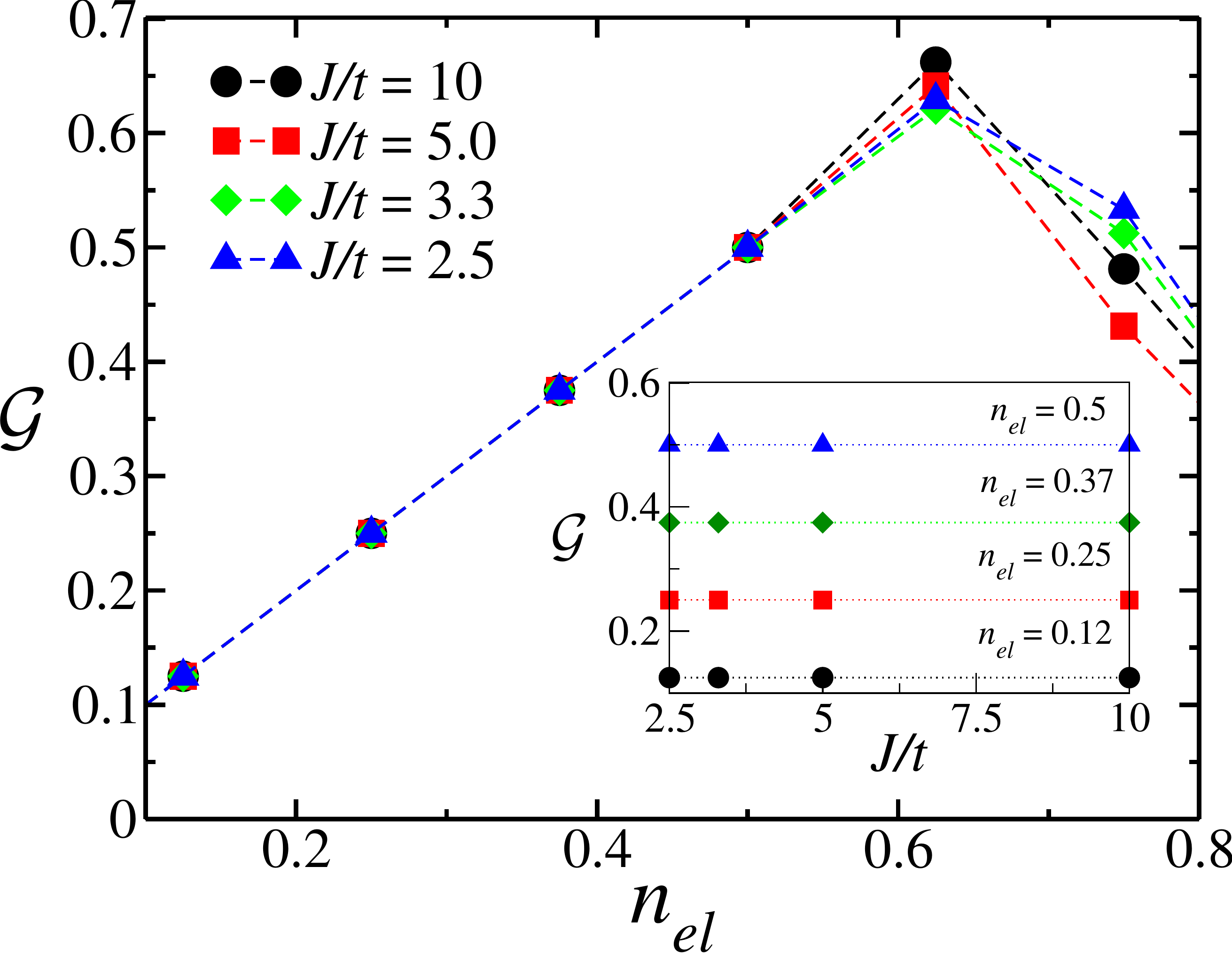}
\caption{(Color online) Adiabatic freezing of genuine multipartite entanglement. 
The plot shows the variation of 
the generalized geometric measure, 
$\mathcal{G}$, with $n_{el}$ for different values of $J/t$.  
The number of lattice sites in the 1D model is fixed at $N = 16$. At low electron density, viz. $n_{el} \lesssim 0.5$, $\mathcal{G}$ 
increases linearly, \emph{along the same line}, with $n_{el}$, and reaches its maximum value
at $n_{el}\approx 0.6$. This feature remains invariant for any value of the $J/t$ ratio. However at large $n_{el}$, $\mathcal{G}$ 
becomes a function of system parameters and the feature -- of increasing along the same line -- obtained earlier,  disappears. 
The inset shows that  $\mathcal{G}$ is
frozen with respect to change in $J/t$, for low $n_{el}$. 
The axes dimensions are the same as in Fig.~\ref{fig1}.
}
\label{ggm}
\end{figure}


We study the variation of GGM in the ground state of the 1D $t$-$J$ Hamiltonian, with respect to system parameters $J/t$ and $n_{el}$, as depicted in Fig.~\ref{ggm}. For convenience in representation, we look at higher values of $J/t$ ($\geq 2.5$), corresponding to the superconducting and PS phases of the model. 
We observe that $\mathcal{G}$ increases linearly with $n_{el}$, at low 
values of $n_{el}$, for fixed $J/t$. It reaches a maximum at
$n_{el} \approx$ 0.6, thereafter decreasing with further increase in $n_{el}$. This behavior is similar to the ground state properties of spin liquid phases 
in doped Heisenberg ladders \cite{t-J_RVB_ours}. 
Significantly, in the low electron density regime, i.e., $n_{el} \lesssim$ 0.5, the genuine multisite entanglement ($\mathcal{G}$) 
is {insensitive} to the parameter $J/t$, and is thus adiabatically frozen. We have numerically observed that at low $n_{el}$ this phenomenon extends to lower values of $J/t$.
However, this freezing of GGM completely vanishes as the electron density is increased. 
 We note that such adiabatic freezing of ME is not observed in other models, for instance in the undoped anisotropic 1D model  \cite{diverging-roy}.
 

This highlights a set of very unique features of the ground state phases of the  1D $t$-$J$ Hamiltonian.
In particular, in the metallic Luttinger liquid phase, at low $J/t$ and  $n_{el}$, bipartite entanglement is long-ranged and adiabatically frozen, in stark contrast to the exponentially decaying BE in superconducting and PS phases. However, at low $n_{el}$ but all $J/t$, including the latter phases, multipartite entanglement is frozen and completely invariant to system parameters. 
This provides an interesting interplay between the behavior of BE and ME in different phases of the doped Hubbard model.

\section{\label{conc}Conclusion}
Entanglement is an important resource in 
quantum information protocols \cite{amico,adv,horo}. However, in general, both bipartite and multipartite entanglement are 
fragile to decoherence \cite{sudden_death}, and this is one of the main obstacles in realization of these protocols. 
Moreover, entanglement may also be highly sensitive to perturbative or sudden 
changes in system parameters and may fluctuate close to critical points, as observed during collapse and revival \cite{hsd-epl-pla}
and dynamical transitions of entanglement \cite{aditi-pra}. 
It was observed that 
certain information-theoretic quantum correlations,
such as quantum discord, could exhibit freezing in the face of decoherence \cite{freezing}, espousing a strong belief that this could 
lead to robust information protocols. However, entanglement,  the workhorse of key quantum information protocols,
rarely freezes under system parameter or temporal changes, including under decoherence 
(cf.~\cite{ekhane-pinjar}). Our results show that 
doped quantum spin chains described by the 1D $t$-$J$ Hamiltonian contain ground state phases that exhibit adiabatic freezing of both bipartite and genuine 
multisite entanglement. Interestingly, the same model without the insertion of defects -- in the form of doping -- does not exhibit a similar 
freezing phenomenon \cite{diverging-roy}. 
It is the presence of defects 
in the quantum spin system that gives rise to the nascent phenomenon of adiabatic freezing of entanglement. 
An important observation in this regard is that no freezing phenomenon of multiparty entanglement (or other multiparty quantum correlations) has hitherto been observed in any quantum system. 
For applications in quantum information protocols, such as fault-tolerant \cite{kit} or one-way computation \cite{comp}, robustness of 
multisite entanglement over fluctuating system parameters can be a significant resource in achieving desired levels of stability.

\acknowledgments
The research of SSR was supported in part by the INFOSYS scholarship for senior students.
HSD acknowledges funding by the Austrian Science Fund (FWF), project no. M 2022-N27, under the Lise Meitner programme. DR acknowledges support from the EU Horizon 2020-FET QUIC 641122.

\end{document}